\documentclass[12pt]{iopart}

\usepackage{harvard}
\usepackage{iopams}
\usepackage{dcolumn}  
\newcolumntype{f}[1]{D{.}{.}{#1}}

\usepackage{epsfig}

\begin{document}

\title[Photon emission from highly charged heavy ions]{Relativistic
       quantum dynamics in strong fields:
       Photon emission from heavy, few-electron ions}

\author{S Fritzsche$^{\,1}$, P Indelicato$^{\,2}$ and Th St\"o{}hlker$^{\,3}$}

\address{$^{\,1}$\  Institut f\"u{}r Physik, Universit\" at Kassel,
                    Heinrich-Plett-Str.\ 40, D-34132 Kassel, Germany}
\address{$^{\,2}$\  Laboratoire Kastler Brossel,
                    \'Ecole Normale Sup{\'e}rieure
                    et Universit{\'e} Pierre et Marie Curie,
                    Boite 74, 4 Place Jussieu, F-75252 Paris CEDEX 05, France}
\address{$^{\,3}$\  Gesellschaft f\"ur Schwerionenforschung (GSI),
                    D-64291 Darmstadt, Germany,
                    Institut f\"ur Kernphysik University of Frankfurt,
            60486 Frankfurt, Germany}

\begin{abstract}
Recent progress in the study of the photon emission from
highly-charged heavy ions is reviewed. These investigations show
that high-$Z$ ions provide a unique tool for improving the
understanding of the electron-electron and electron-photon
interac\-tion in the presence of strong fields. Apart from the
bound-state transitions, which are accurately described in the
framework of Quantum Electrodynamics, much information has been
obtained also from the radiative capture of (quasi-) free
electrons by high-$Z$ ions. Many features in the observed spectra
hereby confirm the inherently \textit{relativistic} behavior of
even the simplest compound quantum systems in Nature.
\end{abstract}


\maketitle

\section{Introduction}
\label{introduction}


Hundred years after Einstein put forth his ideas about relativity and the particle
nature of light, the photon emission from highly-charged, heavy ions has been found
a unique and very exciting framework  for studying his visionary concepts in detail.
The gradual discovery that only the combination of relativity with the photon
picture in the framework of quantum mechanics (the theory that was needed to
understand the microscopic world) could describe the interaction between light
and matter in all its diversity is well exemplified by recent case studies on
high-$Z$ ions. This combination has lead also to Quantum
Electrodynamics (QED), the most accurate and successful theory in
physics today, and served as model for those Field Theories that
now compose the Standard Model. \textit{Relativistic}
transformations are also required to interpret the experiments
with high-energetic ions, if their speed becomes a sizeable
fraction of the speed of light. Indeed, experiments using Laser
spectroscopy at ions storage rings have enabled to test with high
accuracy time dilation \cite{skeh2003}. Born less than 3 decades
ago, the field of highly charged ions is therefore a tribute to
Einstein's work.

Today, there are two research lines for which the observed photon
spectra are crucial for our present understanding of the
light-matter interaction in the presence of strong fields. They
are related to the electronic \textit{structure} and
\textit{dynamics} of high-$Z$ ions and to great improvements in
the accuracy of the experiments. Our (theoretical) understanding
of atoms and ions has advanced considerably during the last decade
thanks, for example, to the study of x-ray transitions between
bound states of high-$Z$ ions. These investigations have
established heavy few-electron ions as a privileged tool owing to
the strong enhancement of QED and other relativistic effects by
large powers in $(Z\alpha)$.

In relativistic ion-atom collisions, in addition, much details about the
electron-photon inter\-action in strong fields were obtained from the
radiative capture of free (or quasi-free) electrons (REC).
At storage rings, this electron capture
determines not only the life\-times but also provides information about the
magnetic components of the radiation field and the coupling of the spin and
orbital motion of the electrons. Recent experimental and theoretical advances
showed, furthermore, that REC may provide a tool for controlling the
polarization of ions beams.

In this contribution, the recent progress in the study of the photon emission
from highly-charged heavy ions is reviewed. After a short summary on the
experimental facilities for heavy-ion research in Sec.~\ref{experiment}, the
photon emission from bound-bound transitions is discussed in Sec.~\ref{sec:rel-qed},
including the Quantum Electrodynamic corrections to the transition energies
as well as the one- and two-photon emission from high-$Z$ ions. In
Sec.~\ref{sec:ecapt}, we review the radiative capture of free electrons with
some emphasis on the polarization of the emitted radiation and the alignment
of the residual ions, if the capture occurs into an excited state of the ion.
Conclusions and outlook onto future experiments are finally given in
Sec.~\ref{outlook}.

\section{Experimental heavy ion facilities}
\label{experiment}

\begin{figure}
\centerline {
\epsfig{file=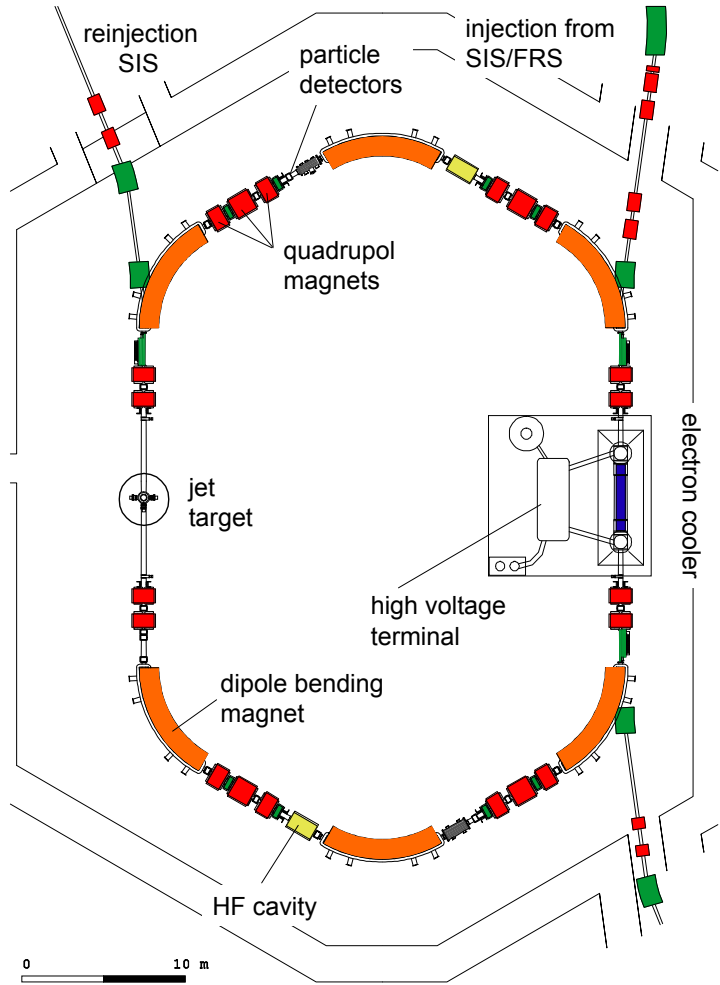,bbllx=3pt,bblly=54pt,
    bburx=576pt,bbury=811pt,height=14.cm,clip=}
} \caption{Schematic presentation of the storage and cooler ring
ESR at GSI-Darmstadt. The layout depicts the beam guiding system
(dipole bending magnets, quadrupoles and hexapoles) as well as the
most important installations for beam handling and diagnostics
(kicker, rf cavities, Schottky noise pick up, electron cooler).
The position of the internal jet-target is marked in addition.
} \label{fig:storage-ring}
\end{figure}

The current progress in the basic fields of atomic collision- and
structure-research involving highly charged heavy ions is closely
related to the application of modern ion source and accelerator
techniques as well as to the use of advanced detection techniques
for photons, electrons and recoil ions. During the last few years
the development of storage rings equipped with electron-cooler
devices \cite{Franzke1987,Poth1990,Bosch1993,Mokler1996,Steck2004}
and electron beam driven ion traps have received a lot of
attention \cite{Schneider1989,Marrs1994a,Marrs1994b,Gillaspy2001}.
For the heaviest ions such as hydrogenlike uranium, a quantum leap
was achieved with the advent of the heavy-ion storage ring ESR at
GSI in Darmstadt (see Fig. \ref{fig:storage-ring}) and the
Super-EBIT at Livermore. At the ESR, electron cooling guarantees
for ion beams of unprecedented quality, i.e.\ this technique
provides cooled and intense beams at high-$Z$ and with precisely
known energies and charge states at small momentum spread
\cite{Bosch1993,Mokler1996,Steck2004}. These conditions are in
particular well suited for the spectroscopy of x-ray transitions
in the heaviest H-like ions. In contrast to storage rings, at EBIT
devices, the highly charged ions are produced at rest in the
laboratory. There, the experiments focus on QED and atomic
structure studies for heavy few-electron ions
\cite{bkme93,boec95,Gillaspy2001}.

In the following, we concentrate on the experimental techniques at
the heavy-ion storage ring ESR where  radiative recombination and
REC transitions have become a subject of detailed experimental
investigations. At the ESR, interaction of the ion beams with
matter can be studied under single collision conditions at the
internal gasjet target where particle densities of about
10$^{12}$p/cm$^3$ are provided. This can be compared with a
typical density of a solid state target of about
10$^{21}$p/cm$^3$. Most important and in contrast to conventional
single-pass experiments where direct beams from relativistic
accelerators are used, no active or passive beam collimation is
required at the ESR. Thus, experimental conditions are almost
completely background-free. A further unique feature of the ESR is
the deceleration capability of the storage ring. It nenables to
perform atomic collision experiments for highly-charged ions in a
completely new energy and charge-state domain, i.e. for highest
atomic charges (e.g. U$^{92+}$) at energies far below their
production energy \cite{Stoehlker1998,Steck2004}. In this
low-energy domain, the perturbation $Q/v$ ($Q$ and $v$ the charge
and the velocity , respectively) caused by the projectile reaches
values otherwise not accessible at accelerators. Furthermore, the
deceleration technique turned out to be indispensable for accurate
precision spectroscopy aiming for a test of QED at high-$Z$ H-like
ions such as H-like uranium. For low-energetic ion beams the
relativistic Doppler corrections are strongly reduced whereas at
high energies the Doppler effect is a serious limitation for such
studies. As an example, by applying this technique for U$^{92+}$ a
beam energy of 3~MeV/u could already be achieved (corresponding to
a velocity of 8\% of the speed of light $c$) \cite{Steck2004}
which has to be compared with the production energy for the bare
charge state of 400~MeV/u (71\% of $c$).

\begin{figure}
\centerline {
\epsfig{file=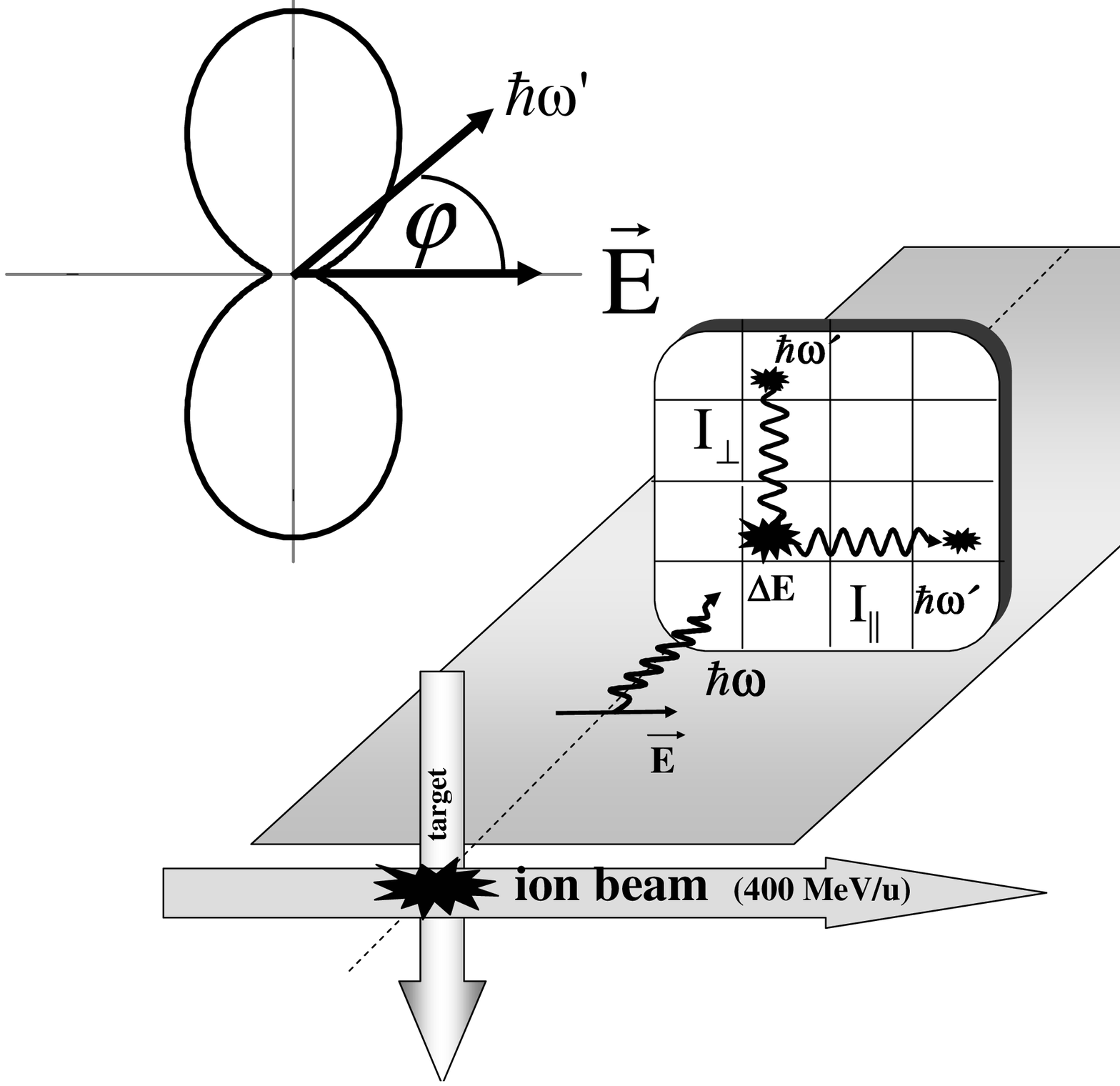,bbllx=93pt,bblly=263pt,
    bburx=788pt,bbury=798pt,height=9.5cm,clip=}
} \caption{Detector geometry used for the measurement of the
linear photon polarization for K-REC at 400 MeV/u
U$^{92+}$$\rightarrow$N$_2$ collisions by exploiting the Compton
effect (see also Sec. ~\ref{sec:pol-kshell})
\cite{Stoehlker2004,Tashenov2004}.
} \label{fig:polarization-detector}
\end{figure}

During the last decade, the progress in storage ring and cooling
techniques was accompanied by an impressive development of
position and energy sensitive solid state detectors for advanced
photon spectroscopy. The tremendous progress in this field of
detector design, which took place very recently, is mainly
motivated by the demands for efficient $\gamma$- and x-ray
spectrometers, in connection with the need for such devices that
arises in applied research such as medical imaging.
The properties of such detectors are millimeter to
sub-millimeter spatial resolution as well as time and energy
resolution in the hard x-ray energy regime above 15~keV
\cite{Protic2001,Stoehlker2003}. Combined with a focusing crystal
spectrometer, for example, these detectors make possible the
measurement of an energy spectrum wide enough to investigate the
whole  energy range of interest simultaneously \cite{Beyer2004}.
Very recently, moreover, a microstrip detector system was
developed at the Forschungszentrum J\"ulich \cite{Protic2001} with
a position resolution of close to 200~$\mu$m which has become
available for the high-precision x-ray spectroscopy at the ESR
storage ring \cite{Stoehlker2003,Beyer2004}. Along with a new kind
of transmission crystal spectrometer \cite{Beyer2004}, such
detectors may play a key role for a precise test of quantum
electrodynamics in the heaviest one-electron systems. Another very
important feature of granular, position sensitive systems is their
sensitivity to the photon polarization at energies above 100~keV.
Using two-dimensional solid-state detectors, it was shown recently
that the polarization of bound-bound and free-bound transitions in
highly-charged heavy ions can be measured with high accuracy by
exploiting the Compton scattering within the detector
\cite{Stoehlker2004}. This is illustrated in
Fig.~\ref{fig:polarization-detector}, where the detection geometry
for a polarization experiment is displayed, using a germanium
pixel detector (see also Sec.~\ref{sec:pol-kshell}).


%
%
%

\section{Relativistic and Quantum Electrodynamics effects in photon emission}
\label{sec:rel-qed}

The photon emission from highly charged ions has many specific
aspects which all related to the fact that the speed of the
electron on its orbit is of order $Z\alpha c$, which is worth
66~\% of the speed of light for the $1s$ shell of uranium. The
relativistic effects enter the transition rates by both the
transition energy and the operator. In this section we summarize a
number of theoretical considerations for both energies and
transition rates and discuss a number of experimental results.

\subsection{Quantum Electrodynamics corrections to transition energies}
\label{subsec:ener-qed}

\begin{figure}
\centerline {
\includegraphics[width=12cm]{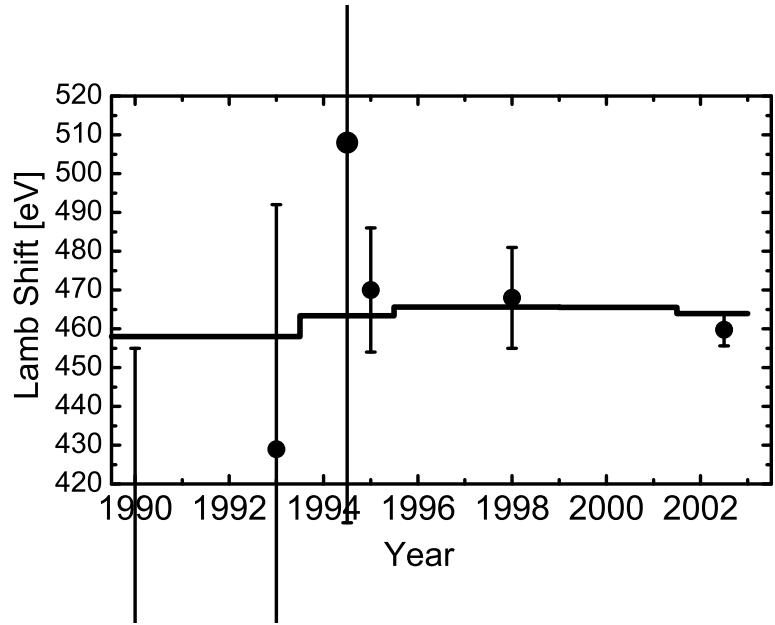}
}
\caption{Evolution of the accuracy of the Lamb shift measurements and
the calculations (solid line) over time for hydrogen-like uranium
[see \citeasnoun{gsbb2004} and Refs.\ therein].
}
\label{fig:ulamb}
\end{figure}

Relativistic Quantum theory really started when \citeasnoun{dir28}
proposed the equation bearing his name. Because he started from
the relation between mass, impulsion and energy, $E^2=m^2c^4+p^2
c^2$, introduced by Einstein's relativity, Dirac found that his
equation had both positive and negative energy solutions. He was
thus lead to the concept of the Dirac \textit{electron sea} to
avoid transitions from positive energy states to negative energy
ones.  Soon, moreover, the relativistic form of the
electron-electron interaction was investigated by
\citeasnoun{bre29}. The existence of the Dirac sea lead to the
idea of Vacuum Polarization, that had calculable effect on atomic
level energies \cite{ueh35}. Yet the theory was plagued by
infinities showing up in all sorts of perturbation expansions. The
experiment of  \citeasnoun{lar50} showing that the Dirac equation
could not predict correctly the fine structure of hydrogen was one
of the experiments that lead to QED. After the first evaluation of
the self-energy  by \citeasnoun{bet47}, the Lamb shift was
calculated more and more accurately in the framework of
non-relativistic QED (NRQED) for many years, as a series in
$Z\alpha$. It is only in the 70's that \citeasnoun{moh74b} showed,
by performing the first high-precision, all-order calculation of
the $1s$ Lamb-shift, that the NRQED $Z\alpha$ expansion did not
converge, even for moderately large $Z\ge 10$. Since then a
considerable amount of non-perturbative QED calculations have been
performed for heavy one- and few-electron ions, including a
complete calculations of QED corrections of order  $\alpha^2$,
e.g., two loop self-energy, see \citeasnoun{yis2003b} and
references therein. For a review of non-perturbative, one-electron
QED calculations, see, e.g., \citeasnoun{mps98}. A few years
later, a series of experiments measuring the hydrogen-like ions
$1s$ Lamb shift started, the first one being peformed at the
BEVALAC in Berkeley \cite{bcid90}. Within the past 14 years,
thanks to the use of heavy-ion storage rings, the experimental
accuracy for the $1s$ Lamb-shift has improved by a factor of 25,
although it cannot yet match theoretical accuracy
(Fig.~\ref{fig:ulamb}). The most recent values for the different
QED and nuclear corrections to the $1s$ Lamb shift in
hydrogen-like uranium are presented in Table \ref{tab:ulamb} and
compared with  recent experimental results obtained at ESR.

Following the first experiment on Lithium-like uranium
\cite{sbbc91}, a large effort has also been made in the evaluation
of QED corrections for three-electron systems
\cite{iam91,cjs91,cjs93,yabs98,yabs99,yass2000,iam2001}.
Simultaneously the use of electron storage rings provided many
accurate measurements \cite{bkms2003}. In principle, QED is the
theory of choice to perform such calculations for heavy ions. Yet,
as soon as a high accuracy is required, QED cannot be used alone
in practice, even for relatively large $Z$. This is because the
real behavior of the corrections corresponding to $n$ exchanged
photons between two interacting electrons is only  $1/Z^{n}$ while
a naive look at the corresponding Feynmann diagrams would lead to
expect a dependence in $\alpha^{n}$. One would then have to
evaluate Feynman diagrams of very high order to reach an
acceptable precision, which is  impossible with presently known
QED techniques, although some explorations are being done
\cite{lin2000}.  In practice, therefore, one has to resort to
relativistic many-body methods like Relativistic Many Body
Peturbation Theory (RMBPT), Relativistic Configuration Interaction
(RCI) and Multi-Configuration Dirac-Fock (MCDF), which are
described in Sec.~\ref{subsec:manybody}, and to correct for
missing QED contributions order by order if doable.

\begin{table}
\caption{Contributions to the $1s$ Lamb shift of Hydrogen-like
uranium \cite{yis2003a}. SE: Self-energy; Uehling: Vacuum
polarization in the Uehling approximation; WK: Wichmann and Kroll
correction to vacuum polarization; Fin.\ Size: effect of the
nuclear charge distribution, assuming a mean spherical radius of
5.860(2)~Fm;  Nucl. Pol.: nuclear polarization \cite{pas95};
two-loop: sum of all QED corrections of order $\alpha^2$.
Experimental value from \cite{gsbb2004}.}
\label{tab:ulamb} \centering
\begin{tabular}{llf{7}}
\\[-0.3cm]  \hline
& Contrib.& \multicolumn{1}{c}{Value (eV)} \\
\hline
$\alpha$ QED & SE & 355.046 \\
& Uehling & -93.597 \\
& WK & 4.975 (2) \\
\hline
$\alpha^2$ QED & Two-loop & -1.26 (33) \\
\hline
& Recoil & 0.46 \\
Nucl. Effects & Fin. Size & 198.79 (40) \\
& Nucl. Pol. & -0.19 (9) \\
\hline
& total& 464.22 (53) \\
& experiment & 460.2 (4.6) \\
\hline
\end{tabular}
\end{table}

\subsection{Relativistic Many-Body issues}
\label{subsec:manybody}

The need for relativistic self-consistent field technique was felt soon after
the intro\-duc\-tion of the Dirac equation \cite{swi35}. At first the
relativistic many-body techniques were developed from their non-relativistic
counterparts, replacing the Schr\"odin\-ger operator by the Dirac one, and
replacing the Coulomb interaction between the electrons by the
Breit interaction.

Over the years relativistic calculations evolved until the
development of the MCDF method by \citeasnoun{gra70} and
\citeasnoun{des75}. Because this method is very general and
(easily) provides a large fraction of the many-body contributions,
the so-called correlation energy, it became rapidly popular. Yet
its ties to QED and the role of the negative energy states,
inherent to the use of the Dirac equation, was not considered for
some years. The existence of negative energy states lead to a
problem known as \emph{continuum dissolution} \cite{bar51,suc80}:
treatment to all orders using an electron-electron interaction
operator that couple positive and negative energy states (which is
the case of even the Coulomb interaction) lead to infinities. No
rigorous solution to this problem can be found outside of QED.
Indeed, derivation of the many-electron Hamiltonian from QED shows
that the electron-electron interaction \emph{must} be
``sandwitched'' between projection operators on the positive
energy states.  It is only recently that the push toward high-$Z$
few-electron ion experiments prompted a study on how continuum
dissolution happens in MCDF calculation and how to implement
projection operators \cite{ind95}.

Relativistic Many-Body Perturbation Theory evolved originally from the
non-relativistic work by \citeasnoun{kel63} and \citeasnoun{lin74}.
Early calculations were performed on lithium-like ions \cite{jbs88}.
A review on RMBPT techniques and results can be found in \cite{sap98}.
Hereby, the implementation  of projection operators in RMBPT was very
natural and mandatory as infinities appear already in second order
perturbation theory if such operators are not present
\cite{hlll86,hllm86}.

An other method that has been used in few-electron calculations is
the Relativistic Configuration Method (RCI). This method is a
variant of the MCDF method, which uses finite basis sets like
RMBPT. It was found to be very effective to describe few-electron
ions \cite{ccj93,ccjs94,jsc95,ccj2001}. MCDF and RCI calculations
have been tested in experiments realized at the Super-EBIT in
Livermore, on transition to the $n=2$ levels of 3 to 10-electron U
and Th ions\cite{bkme93,boec95}.

Beside the need for projection operators, there is a number of features
that are common to all relativistic many-body calculations.
First, the electron-electron interaction contains retardation terms,
which is a direct consequence of Einstein's relativistic theory
(finite speed of light). Because each particle in the system must have
its own proper time in relativity, there can be no exact Hamiltonian
formalism to describe it. This can be seen, for instance, in the expression of
 the electron-electron interaction
operator, which depends on the one-electron orbital energies. These energies are knwon only when the full solution
of the self-consistent field has been found and cannot be defined \textit{a priori}
in the many-body formalism.

Because of the multitime nature of the problem, the most general
formalism to handle the relativistic many-body problem is the
two-time Green's function formalism of \citeasnoun{sha90}. This
formalism is needed, in particular, to add QED corrections to the
transition rates and other quantities beyond the energies. An
alternative formalism has been developed recently \cite{lsa2004},
based on covariant evolution operator. At some point, however, QED
corrections to second order in the electron-electron interaction
must be made \cite{bmjs93,lpsl95,mas2000,assl2002}. These
corrections partially cancel two-electron self-energy corrections.
At present, this interplay between QED and many-body effects
constitutes the greatest challenge posed to the accurate
theoretical evaluation of transition energies in the field of
highly-charged heavy ions.

\subsection{One-photon bound-bound radiative transition}

\begin{figure}
\centerline {
\includegraphics[width=8cm,clip]{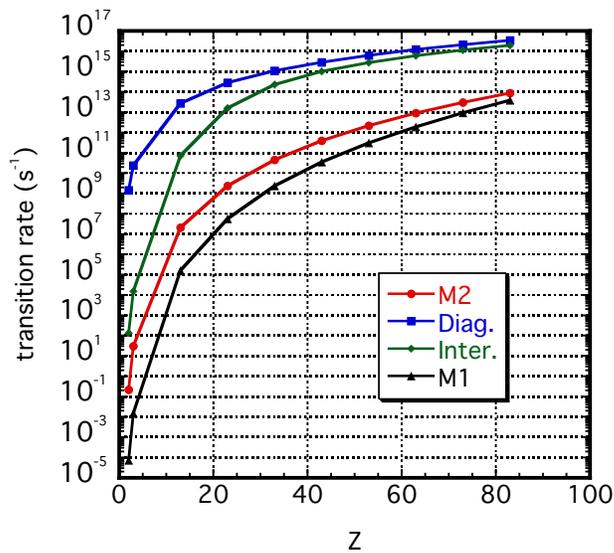}
} \caption{Transition rate for He-like ions (s$^{-1}$). M$_{2}$:
$1s2p\,^{3}P_{2}\to 1s^{2} \,^{1}S_{0}$. Diag: diagram line
$1s2p\,^{1}P_{1}\to 1s^{2} \,^{1}S_{0}$ (E$_{1}$). Inter:
intercombination line $1s2p\,^{3}P_{1}\to 1s^{2} \,^{1}S_{0}$
(E$_{1}$). M$_{1}$: $1s2s\,^{3}S_{1}\to 1s^{2} \,^{1}S_{0}$.
} \label{fig:helife}
\end{figure}

Relativistic effects play a central role also in the photon
emission. In the helium isoelectronic sequence, for example, the
$1s2s\;^{3}S_{1}\to 1s^{2} \;^{1}S_{0}$ line is called the
``relativistic M$_{1}$'' transition because it is completely
forbidden in non-relativistic theory. For high-$Z$ ions , in
addition, intercombination lines like $1s2p\;^{3}P_{1}\to 1s^{2}
\;^{1}S_{0}$ become almost as intense as the ``allowed''
transitions, i.e.\ the diagram line $1s2p\;^{1}P_{1}\to 1s^{2}
\;^{1}S_{0}$. The dependence of the transition rates as a function
of the atomic number and the successive multipoles, obtained by
the expansion of the relativistic transition operator in spherical
components, is such that high-multipole transitions  can occur
with sizeable probabilities. Compared to its non-relativistic
equivalents, this operator automatically includes such
relativistic effects as retardation. To stay with the example of
the helium-like ions, the $1s2p\;^{3}P_{2}\to 1s^{2} \;^{1}S_{0}$
M$_{2}$ transition has a relative strength of $1.5\times 10^{-10}$
at $Z=2$, but already $2.5\times 10^{-3}$ at $Z=83$. Some example
of the evolution of transition probabilities in helium-like ions
are shown on Fig.~\ref{fig:helife} and more details can be found
in \citeasnoun{mam78}. The lifetime of the $1s2s\;^{3}S_{1}$ in
He-like Xe has been measured at GANIL to 3~\% accuracy
\cite{mcib89} and has been found in good agreement with theory
\cite{jps95,ind96}.

In general, the higher-order multipole contributions to a given transition
rate can be neglected. However, there are a few cases for which this is not
true. For example, in the $3d_{5/2}\to 2p_{1/2}$ transition in hydrogen-like
ions, the M$_{3}$ contribution represents between 32~\% ($Z=1$)
and 40~\% ($Z=92$) of the dominant E$_{2}$ multipole. Similarly, the
E$_{4}$ multipole contributes between 51~\% to 58~\% of the
$4f_{7/2}\to 2p_{1/2}$  M$_{3}$ transition for these ions. The study of
highly-charged heavy ions has enabled one to observe directly
forbidden transitions beyond M$_{2}$ transitions, which has been observed
or relatively low-$Z$ for helium-like ions. For example, magnetic octupole
(M$_{3}$) transitions have been observed in an EBIT for nickel-like Th and U
ions \cite{bosw91}. Even if the contribution of the higher multipoles
to the total rate is of the order of 1~\%, they can however influence
the angular distribution of the emitted photons. Recent experiments on
that subject are described in Sec.\ \ref{sec:em-char-rad}.

Transition probabilities in few-electron ions are a very stringent
test of relativistic many-body theories. In variational methods
like the MCDF method, transition probabilities involve the
wavefunction, which is less precisely evaluated than the energy.
In the same way as in the evaluation of transition energies, there
are several issues that have to be addressed when calculating
transitions probabilities. For example, one must also properly
account for the negative-energy continuum. This was first noticed
for the $1s2s\,^{3}S_{1}\to 1s^{2} \,^{1}S_{0}$ M$_{1}$ transition
\cite{ind96}. It was also shown, that in contrast to the
non-relativistic case, the full gauge invariance (for example
between length and velocity gauge for E$_{n}$ transitions) can be
achieved \emph{only} if the negative energy continuum is properly
accounted for \cite{dsij98}. An extra complication, when using
highly correlated wavefunctions for evaluating transition
probabilities and other operators, is that the orbitals in initial
and final wave functions are usually not orthogonal \cite{caj77}.
This may have a large effect in some transitions like the
relativistic  M$_{1}$ \cite{ind96}. Other kinds of transitions
like two-electron, one-photon transitions, may depend entirely on
non-orthogonality between correlated wave functions. A competition
between such a transition and an E$_{2}$ transition has been
predicted and observed in Be-like xenon \cite{ind97}.

There has been little studies of QED corrections to transition
amplitudes. Most of the time, QED corrections are accounted for
only by the inclusion of radiative corrections to the energy
(which is done automatically when experimental energy are used for
the calculation of the transition rates). QED corrections to the
$2p\to 1s$ and $2s\to 1s$ in hydrogenlike ions have been
calculated recently \cite{spc2004}. These authors have shown that
there might be a very strong cancellation between the effect of
the energy correction and those of the wavefunction correction.
When several electrons are present, there are other kind of QED
corrections that needs to be included. For example, the reducible
contribution to some transitions in helium-like ions has been
investigated \cite{isv2004} and found to be small.

Forbidden transitions can also be a tool for studying the
interaction of the electrons with the nuclear magnetic moments
through the hyperfine quenching of the $1s2p\,^{3}P_{0}$ level in
heliumlike ions which was calculated by the MCDF method
\cite{ipm89} and later by RMBPT \cite{jcp97}. This effect was then
observed in a variety of ions from nickel to gold
\cite{msid89,dllb91,ibbc92,bbcd93,tmbb2004}. Even the nuclear
quadrupole moment can have a large effect for the lifetimes of
metastable states (see, e.g.\ \cite{pmi94}). The most recent
experiment \cite{tmbb2004}, is the  heir of the thirty-years old
Beam-Foil technique. Yet its accuracy is the result of several
factors. It uses the increase
 in the lifetime of the ion due to relativity at high-beam energies. The set-up comprises
 a magnetic spectrometer and an advanced
position-sensitive beam detector to detect the photons from the
metastable state in coincidence with ions of the associated
charge-state. Finally it benefits from  the high quality of
electron-cooled ions beam from the SIS synchrotron in GSI.

\subsection{Two-photon emission}

\begin{figure}
\centerline { \epsfig{file=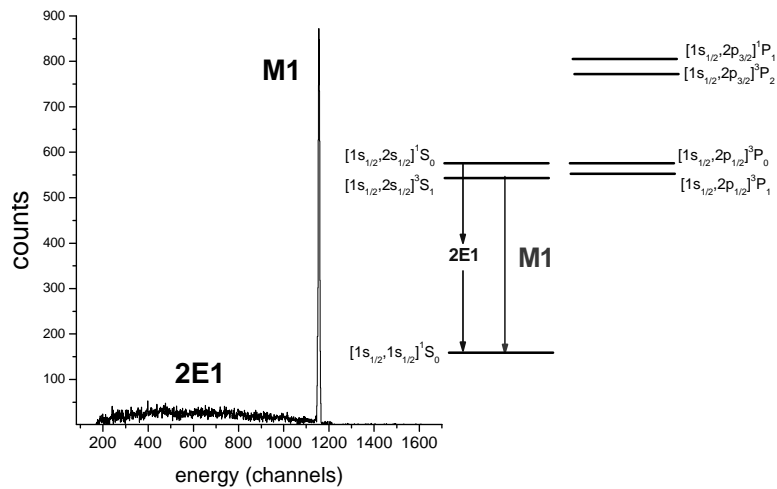,bbllx=100pt,bblly=307pt,
    bburx=750pt,bbury=720pt,height=6.5cm,clip=}}
\caption{X-ray spectrum observed for Li-like uranium U$^{89+}$ in
collisions with N$_2$,  measured in coincidence with the
projectile ionization. i.e. U$^{90+}$. The spectrum is entirely
governed by a single K$\alpha$ transition stemming from the M$_1$
decay of the $1s2s\,^3S_1$ state. The broad feature is due to the
two-photon decay (2E$_1$) of the $1s2s\,^1S_0$ level. }
\label{fig:2E1M1}
\end{figure}

Two-photon transitions can become important in heavy ions when a
level cannot decay by other ways, e.g., when only strictly
forbidden $J=0 \to J=0$ transitions would be possible otherwise.
The  2E$_{1}$ two-photon transition, for instance, dominate  the
lifetimes of the $2s$ and $1s2s\;^{3}S_{1}$ levels at low-$Z$, but
not at high-$Z$, because of the strong $Z$-dependence of the
M$_{1}$ transition. In the absence of nuclear magnetic moments,
the $1s2s\;^{1}S_{0}$ level decays mostly by a two-photon
transition to the $1s^{2}\, ^{1}S_{0}$ ground-state level. For
such two-photon transitions, however, probabilities are difficult
to calculate and to measure since the two-photon spectrum is
spread between zero and the energy difference between the initial
and final levels. Early experiments looked at the single-photon
spectrum, which requires a very good signal to noise ratio, and a
large effort to reduce contaminant X-rays from the environing
material. The heaviest highly charged ions studied by this method
was helium-like Kr \cite{mscb86}. More recently a coincidence
technique was applied where the two emitted photon are detected by
two different detectors. Requiring that the sum of the two-photon
energy is equal to the transition energy, the spectral shape of
the continuum distribution of the 2E$_{1}$ decay was analyzed and
provided accurate results for heavy elements \cite{Schaeffer1999}.
From the theoretical point of view, the transition probability of
two-photon transitions is calculated by second-order perturbation
theory. Both, the positive and negative energy states must be
included in order to have a good correspondence between the
different gauges. The relativistic rates and photon distribution
shapes for the $2s\to 1s$ transition in hydrogenlike ions have
been recalculated recently, taking into account QED correction to
the energy and the nine first multipole contributions
\cite{spi98}. As in the one-photon case, the E$_1$M$_2$ correction
to the dominant 2E$_1$ ranges from a relative contribution of
$3\times 10^{-11}$ at $Z=1$ to 0.2~\% at Z=92. The two-photon
decay rates of several helium-like levels $1s2s\; ^{1}S_{0}$,
$1s2s\; ^{3}S_{1}$ \cite{daj97} and $1s2p\; ^{3}P_{0}$
\cite{saj2002} have been evaluated in a relativistic framework. In
the latter work, it was found that the negative energy continuum
contributes significantly to the the E$_{1}$M$_{1}$ rate from the
$1s2p\; ^{3}P_{0}$ to the ground state. It should be noted that
this two-photon contribution, in competition with the E$_1$
transition $1s2p\; ^{3}P_{0} \to 1s2s\; ^{3}S_{1}$, represents
47~\% of the  $1s2p\; ^{3}P_{0}$ lifetime at $Z=92$. Detailed
measurements of the shape of the two-photon spectrum for different
multipolarities and their comparison with the recent calculations
quoted above remain to be done.

An example for a two-photon spectrum is shown in figure
\ref{fig:2E1M1} in which two  decay modes, the 2E$_1$ transition
from the $1s2s\; ^{1}S_{0}$ state  and the M$_1$ decay of the
$1s2s\,^3S_1$ level are seen simultaneously. In this figure, the
x-ray spectrum observed for Li-like uranium U$^{89+}$ in
collisions with N$_2$ is displayed which was measured in
coincidence with projectile ionization (U$^{90+}$). The spectrum
is entirely governed by an intense single L$\rightarrow$K
transition and a broad continuum distribution. Because we are
dealing with He-like uranium produced by K-shell ionization of the
Li-like species and initially in the $1s^2 2s$ ground state, the
broad continuum can only be explained by the two-photon (2E$_1$)
decay of the $1s2s\,^1S_0$ level while the single K$\alpha$ line
arises exclusively from the M$_1$ decay of the $1s2s\,^3S_1$
state. To the best of our knowledge, no other ion-atom collisions
is known which produces inner-shell excited states with such a
high state selectivity. This unexpected result is currently the
subject of detailed theoretical investigations.

\begin{figure}
\centerline {
\epsfig{file=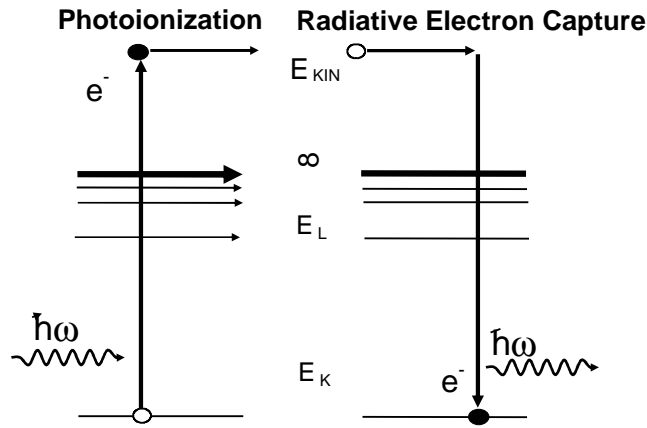,bbllx=136pt,bblly=350pt,
         bburx=796pt,bbury=692pt,height=6.2cm,clip=}
}

\caption{Radiative Electron Capture can be viewed as the time-reversed
photo\-ionization which results in an electron capture into a bound
state of the ion via simultaneous emission of a photon.
}
\label{fig:rec-scheme}
\end{figure}

\section{Electron capture into highly-charged ions}
\label{sec:ecapt}

Apart from the bound-state energies and transitions, strong
relativistic effects become visible also in collisions of
high-$Z$, few-electron ions with electrons and low-$Z$ target
atoms. In these collisions, the electromagnetic field of the
fast-moving projectiles often causes an ionization or capture of
electrons. In particular, the radiative electron capture (REC)
into bare and hydrogen-like high-$Z$ ions has been found to
provide a unique tool for studying the electron-photon interaction
in the presence of strong fields.

\subsection{Radiative electron capture}

REC into highly-charged ions has been investigated since a long
time as it represents the dominant charge exchange process for
bare and H-like ions in collisions with low-Z targets at high
energies
\cite{Schnopper1972,Spindler1979,Anholt1984,Stoehlker1992,Stoehlker1995,Vane2000}.
In this radiative recombination (RR) process, a free or quasi-free
electron is captured into a bound state of the ion under the
simultaneous emission of a photon. In fact, the radiative
recombination of (heavy) ions with free electrons is known also as
the time-reversed photo-effect. Besides the total REC cross
sections, which determine the lifetimes of the ion beams at
accelerators and storage rings, a number of angular distribution
\cite{Anholt1984,Stoehlker2001} and, more recently, polarization
measurements have been carried out
\cite{Stoehlker2004,Tashenov2004} and have shown, that the
radiative capture of electrons is a powerful tool for precise
studies on atomic photoionization with high-energy photons in the
strong-field domain. Since the very first observations of REC (RR)
photons \cite{Schnopper1972}, therefore, this process has been
studied intensively for various bare and few-electron ions,
including bare uranium and projectile energies from a few MeV/u up
to the extrem relativistic regime above 100 GeV/u \cite{Vane2000}.

\begin{figure}
\centerline {
\epsfig{file=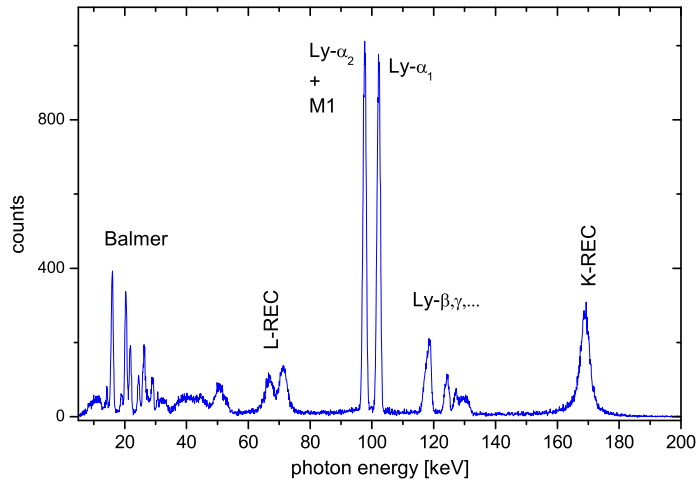,bbllx=0pt,bblly=0pt,
    bburx=307pt,bbury=220pt,height=7.2cm,clip=}
}
\caption{An x-ray spectrum associated with capture for 68 MeV/u
\protect{U$^{92+}$ on N$_2$} is shown. The data were taken at the ESR
storage ring at an observation angle of \protect{$\theta$=132$^o$} (not
corrected for detection efficiency). The x-ray energies in the laboratory
frame are given \cite{Stoehlker1998}.
}
\label{fig:PhotonSpectrum}
\end{figure}

As an example, Figure~\ref{fig:PhotonSpectrum} displays the x-ray
spectrum of H-like uranium which was produced by electron capture
in  U$^{\,92+}$ collision with N$_2$ at 68 MeV/u (recorded at an
observation angle of 132$^{\circ}$).
As in the case of photoionization, the energy of the
REC photons, $\hbar\omega_{\rm\, REC} \,=\, E_{b} \,+\, E_{\rm\,kin}$,
is given by the sum of the binding energy $E_{b}$ and the
kinetic energy $E_{\rm \,kin}$ of the free electron in the
projectile frame (in the present experiment the kinetic electron
energy $E_{\rm \,kin}$ amounts to $\approx$~37~keV). For the REC
transitions into the $1s$ ground state of hydrogen-like uranium
($E_{1s}\approx$~132~keV), the K-REC peak is thus found in
the high-energy part of the spectrum, at a photon energy of around
170~keV. Hereby, the broadening of these K-REC lines (compared to the
characteristic transitions) is due to the momentum distribution of
the target electrons (Compton profile).

Because of the relatively low velocity of the decelerated ions,
the two $j=1/2$ and $j=3/2$ fine-structure components of the L-REC
x-ray lines around 50~keV are still resolved and are separated by
$\sim 4.5$~keV. This illustrates one of the benefits of using
deceleration techniques in REC experiments, while the large line
broadening at high projectile energies, caused by the Compton profile,
often prevents the separation of the fine structure components
\cite{Stoehlker1998}.
The L-shell fine structure splitting, in addition, also leads to
an energy separation of the two Lyman-$\alpha$ ground-state
transitions (Ly-$\alpha_2$ + M$_1$: $2p_{1/2}$, $2s_{1/2}
\rightarrow 1s_{1/2}$, and Ly-$\alpha_1$: $2p_{3/2} \rightarrow
1s_{1/2}$) which constitute the most intense x-ray lines in the
spectrum.

\begin{figure}
\vspace*{1.0cm} \centerline {
\epsfig{file=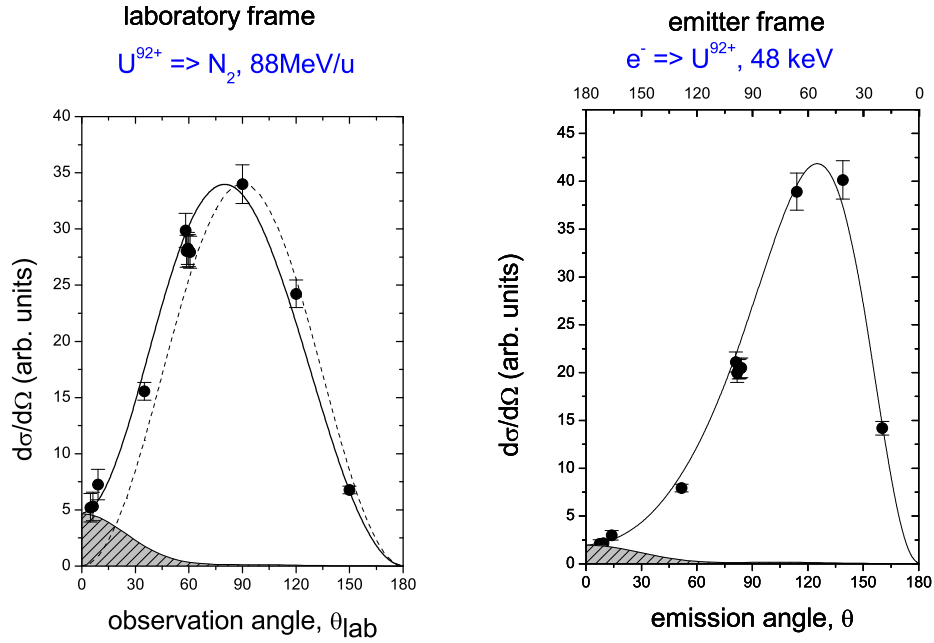,bbllx=0pt,bblly=0pt,
    bburx=310pt,bbury=210pt,height=9.2cm,clip=}
}

\caption{left side: Angular distributions for K-REC at 88 MeV/u
U$^{92+} \rightarrow$N$_2$ collisions \cite{Stoehlker2001}. Solid
circles: experimental result; solid line: relativistic
calculations; shaded area: spin-flip contributions. Right side:
K-REC distribution (solid circles) in the emitter frame as a
function of the emission angle $\theta$ (bottom axis). The
horizontal axis at the top  refers to the corresponding electron
emission angle in photoionization of U$^{91+}$ (photoelectron
energy: 48~keV).
}
\label{fig:AngularDistribution}
\end{figure}

On the left side of Fig.~\ref{fig:AngularDistribution}, the K-REC
angular distribution for the capture into 88~MeV/u U$^{\,92+}$
ions is shown as function of the photon emission angle in the
laboratory frame (solid circles) and are compared with rigorous
relativistic calculations \cite{Stoehlker2001}. As for the
structure calculations in Sec.~\ref{sec:rel-qed}, a relativistic
treat\-ment of the electrons and the electron-photon interaction is
typically required in order to under\-stand the observed data from the
collision experiments \cite{Ichihara1994,Eichler:95,Ichihara1996}.
As seen from the figure, moreover, the measured angular
distribution confirms well the slight asymmetry with respect to a
perpendicular photon emission as predicted by theory. Most
remarkable, however, is the non-vanishing cross section close to
0$^{\circ}$, which demonstrates that the magnetic contributions
are still present in the low-energy domain. Since the magnetic
multipoles contribute 3~\% to the total K-REC cross-section (cf.\
the dashed area in Fig.~\ref{fig:AngularDistribution}),  this
enhancement of the photon emission in forward direction shows the
sensitivity of the applied method. Note that the (almost)
symmetrical angular distribution with respect to 90$^\circ$ is a
particular feature of the laboratory frame which arises from the
cancellation of the various effects due to the retardation of the
electron-photon interaction and the Lorentz transformation for
going from the projectile to the laboratory framework, a behaviour
which was predicted already by the non-relativistic theory
\cite{Spindler1979,Anholt1984}.

In the emitter frame, in contrast, a strong variation occurs for
the angular distri\-bution of the emitted photons as function of
the projectile energies. This is illustrated in
Fig.~\ref{fig:AngularDistribution}(right side) where the observed
data are Lorentz-transformed into the projectile frame. Even for
the low energy regime where the kinetic electron energy $E_{\rm
\,kin}$ is much smaller than the binding energy in the final state
$E_{b}$, that is for a photoionization close to the threshold
\cite{Stoehlker2001}, the angular distribution still exhibits a
considerable backward peaking in accordance with the enhanced
forward emission in the direct photoionization process. This
behaviour of the angular distribution can be understood easily by
replacing $\theta'$ by $\pi-\theta'$ as indicated on the upper
abscissa of Fig\-ure~\ref{fig:AngularDistribution}. Theoretically,
the radiative electron capture and all the subsequent emission
processes are most easily described by means of the density matrix
theory where, instead of a single collision event, an ensemble of
(equally prepared) systems is considered. As appropriate for
collision processes, these systems can be either in a
\textit{pure} quantum state or in a \textit{mixture} of different
states with any degree of coherence \cite{Blum:81,Balashov:00}.
Density matrix theory helps to accompany such collision ensembles
through one or several regions of the interaction without loosing
important information about the reaction products. To support
detailed collision studies, the concepts of the density matrix has
been implemented meanwhile into a number of codes which are
suitable for both, one- and few-electron ions
\cite{Fritzsche:01,Surzhykov:04b}.

\subsection{Polarization studies for the K-shell capture}
\label{sec:pol-kshell}

\begin{figure}
\centerline {
\epsfig{file=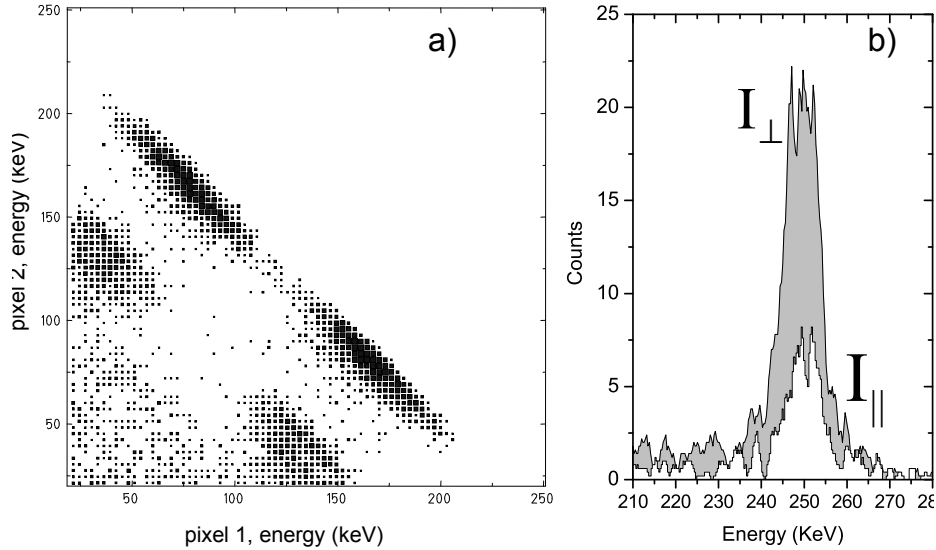,bbllx=80pt,bblly=346pt,bburx=763pt,
    bbury=792pt,height=8cm,clip=}
} \caption{a) Scatter plot of coincident Compton events; b) the
coincident sum energy spectrum for scattering parallel
(I$_\parallel$, white area) and perpendicular (I$_\perp$, shaded
area) to the scattering plane \cite{Stoehlker2003,Tashenov2004}.
}
\label{fig:polarization}
\end{figure}

Details about the radiative capture can be derived not only from
the angular distribution of the emitted photons but also from
their polarization \cite{Surzhykov:01,Eichler2002}. In practice,
however, polari\-zation measurements have been hampered in the
past years because of the lack of efficient Compton polarimeters
for photon energies of several ten or even hundred keV. For
high-$Z$ ions with photon energies above 100 keV, it was
demonstrated only recently that the (linear) polarization of the
emitted photons can be measured by means of a new generation of
segmented germanium detectors, which allow for energy as well as
position resolution \cite{Inderhess1996,Stoehlker2003}. A first
series of polarization measurements were performed at GSI using
these detectors \cite{Stoehlker2003,Tashenov2004} and by applying
the dependence of the angle-differential Compton scattering on the
linear polarization of the incoming photons. As predicted
theoretically \cite{Surzhykov:01,Eichler2002}, a strong linear
polarization of the K-REC photons is expected, which decreases in
the forward direction as the energy of the projectiles is
enlarged. For bare uranium ions at 400 MeV/u, for example, the
photon polarization of the K-REC radiation has been analyzed at
the jet-target of the storage ring ESR. For this purpose, a planar
germanium pixel detector was used, mounted at an observation angle
of 90$^\circ$. In the experiment, the photon polarization is
obtained  by  recording  events which occur simultaneously in two
pixels of the detector. While one pixel is used to measure the
Compton recoil electron ($\Delta E$), the other one records the
scattered photon ($\hbar \omega^{'}$). A scatter plot of such
coincident photon events is displayed in
Fig.~\ref{fig:polarization}. The large number of events in the
diagonal corresponds to events with a (constant) energy sum equal
to the K-REC transition, i.e.\ E$_{K-REC}\,=\,\Delta E + \hbar
\omega^{'}$. It is important to mention that, for our initial
energies (E$_{K-REC}$~$\approx$~250~keV), the condition $\Delta E
< \hbar \omega^{'}$ is always fulfilled which allows us also to
identify the segment where scattering takes place. The latter also
explains the two maxima present in the 2D scatter plot. In
Fig.~\ref{fig:polarization}b, we compare the coincident sum energy
spectrum for scattering parallel (I$_\parallel$) and perpendicular
(I$_\perp$) to the reaction plane (defined by the ion beam and the
propagation direction of the K-REC photon). As seen from this
figure, the K-REC radiation appears strongly polarized within the
scattering plane.

Experimentally, the polarization properties of the emitted photons
are usually obtained from the Stokes parameter, i.e., the
intensity ratios of the light measured under different angles with
respect to the reaction plane. For example, the Stokes parameter
$P_1 \,=\, (I_{0^o} \,-\, I_{90^o}) / ( I_{0^o} \,+\, I_{90^o})$,
is obtained from the intensities \textit{parallel} and
\textit{perpendicular} to the scattering plane, while the
parameter $P_2$ follows from a similar intensity ratio, taken at
$\chi \,=\, 45^o$ and $\chi \,=\, 135^o$, respectively. The two
parameters $P_1$ and $P_2$ together describe the (degree and
direction of the) \textit{linear} polarization in the plane
perpendicular to the photon momentum whereas the third parameter
$P_3$ denotes the degree of \textit{circular} polarization.

In the theoretical treatment of electron capture, the Stokes
parameters are closely related to the photon density matrix
if no further information need to be retained for the remaining ions
apart from its level designation. For the capture of
\textit{unpolarized} electrons by bare ions, it was shown recently
\cite{Surzhykov:03,Fritzsche:03}, that only the Stokes parameter
$P_1$ is non-zero (and positive for moderate projectile energies),
while $P_2$ is identically zero. This implies that, for
unpolarized electrons and ions, the polarization of the
recombination photons will always be found within the reaction
plane. For the capture of \textit{polarized} electrons, in
contrast, the Stokes parameter $P_2$ becomes non-zero, in
particular at small forward angles $\theta_{\rm\, RR}$, leaving
$P_1$ unaffected in this case. For the photon
polarization, however, any non-zero $P_2$ parameter results in a
rotation of the polarization ellipse out of the reaction plane.
Owing to the symmetry of the collision system, a similar result is
found if the (unpolarized) electrons are captured by a polarized
ion beam. Therefore, the rotation of the polarization ellipse may
serve as a unique tool for measuring the polarization properties
of ion beams \cite{Surzhykov:04}, a result which has attracted a
lot of recent interest.

\subsection{Emission of characteristic radiation}
\label{sec:em-char-rad}

If the electron is captured not into the ground state of the
ion, but into an excited one, this ion state decays further towards
the ground state under the emission of \textit{characteristic} radiation.
For high-$Z$ ions, in particular, the capture into the $\,2p_{3/2}\,$
level and its subsequent Lyman-$\alpha_1$ decay into the $1s$ ground
state has been explored in great detail. The angular distribution of
this characteristic radiation,
\begin{eqnarray}
\label{W-of-theta}
   W_{\rm\, Ly} (\theta) &=&
   W_o \left( 1 + \beta_{\rm\, exp}\,P_2 (\cos \theta) \right) \, ,
\end{eqnarray}
then allows to derive an (experi\-mental) anisotropy parameter
$\beta_{\rm\, exp}$ as function of the charge and the energy
of the projectiles [cf.\ Fig.~\ref{fig:AngularDistribution}].
When compared to the standard dipole approximation $\beta \,=\,
\mathcal{A}_{2} / 2$, however, deviations of up to 30 \%{} were
found for the anisotropy parameters $\beta_{\rm\, exp}$, where
$\mathcal{A}_{2}$ refers to the alignment of the $2p_{3/2}$ level
following the radiative electron capture
\cite{Stoehlker1997,Eichler1998}. Initially, this large
discrepancy was quite surprising since, even for hydrogen-like
uranium, the dipole approximation to the electron-photon
interaction was known to provide (theoretical) lifetimes with an
accuracy of better or $\sim$ 1 \%{}. A detailed analysis in the
framework of the density matrix theory later showed, however, that the
increase in the observed anisotropy arises entirely from to the
weak M$_2$ branch of the Lyman-$\alpha$ transitions, i.e.\ from the
interferences of the E$_1$ and M$_2$ multipole components
\cite{Surzhykov/PRL:02}. Theoretically, this enhancement is
understood if, instead of the alignment $\mathcal{A}_{2}$ and
anisotropy parameters $\beta$, the two \textit{effective} parameters
\begin{eqnarray*}
\label{3}
   \mathcal{A}_{\,2}^{\rm\, (eff)}     \, = \,
   {\mathcal{A}_2} \cdot f({\rm E}_1, {\rm M}_2); \;\: \hspace*{1.0cm}
   \beta_{\,20}^{\rm\, (eff)} \, = \, \beta_{\,20} \cdot
   f({\rm E}_1, {\rm M}_2) \,
\end{eqnarray*}
are used in Eq.\  (\ref{W-of-theta}), where
\begin{eqnarray*}
\label{4}
   f({\rm E}_1, {\rm M}_2) & \propto&
   \left[ 1 + 2\sqrt{3}\frac{<||{\rm M}_2||>}{<||{\rm E}_1||>} \right]
\end{eqnarray*}
is called the structure function. This structure function purely
depends on the bound-state structure of the ion, while the
alignment parameters $\mathcal{A}_{\,20}$ and $\beta$ are of
dynamical origin and, hence, are determined by the capture
process. The structure function $f({\rm E}_1, {\rm M}_2)$ is
roughly proportional to $~Z^2$ and, therefore, non-negligible
effects of a few percent from the M$_2$ multipole component may
arise even for medium-$Z$ ions.

\begin{figure}
\vspace*{-0.5cm}
\centerline {
\includegraphics[width=8cm]{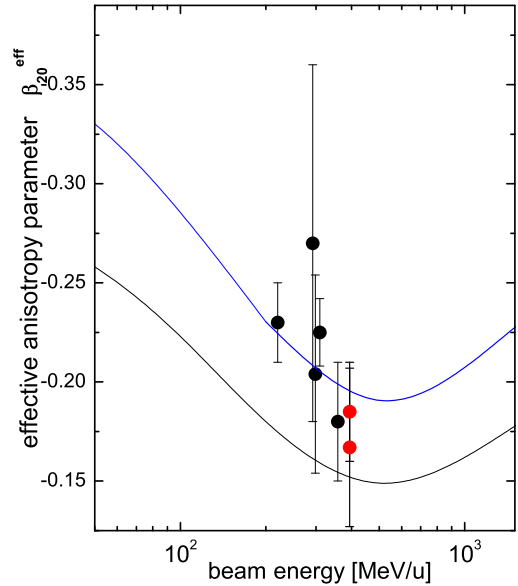}
} \caption{Comparison of experimental and theoretical (effective)
anisotropy para\-meters $\beta$ as function of the projectile
energy for the Lyman-$\alpha_1$ radiation of $U^{\,91+}$ ions,
produced in   U$^{\,92+} \rightarrow$N$_{\,2}$ collisions. The
lower line represents the theoretical prediction from the electric
dipole approximation, while the upper line also includes the
E$_1$-M$_2$ interference \cite{Stoehlker1997,Surzhykov/PRL:02}.
}
\label{fig:alignment-data}
\end{figure}

Apart from the incorporation of higher multipoles, there is an alternative
view of how such \textit{inter\-ference} effects in high-$Z$ ions can be
used to obtain insight into the inte\-raction with the radiation
field. If we assume, for instance, that the REC into bare ions is
well understood by means of the (relativistic) density matrix theory,
we may utilize the theoretical alignment for the capture into the
2p$_{3/2}$ level in order to derive the structure function
$f({\rm E}_1, {\rm M}_2)$ also experimentally. Applied
to the angular distribution data from
Fig.~\ref{fig:alignment-data}, a value $f^{\rm\,(exp)\,} ({\rm E}_1,
{\rm M}_2) \,=\, $1.27$\pm$0.05 is obtained, giving rise to a
relative contribution of the M$_2$ decay branch of $\Gamma_{{\rm
M}_2} / \Gamma_{{\rm E}_1}$ = 0.0077$\pm$0.0009 \cite{Muthig2004}.

\subsection{Multiple photon emission: angular correlations}

A great deal of information  about the electron-photon interaction
in the presence of strong fields has been obtained from the x-ray
spectra of either the REC or the characteristic radiation. Further
details can be derived if the photon emission from the electron
capture and the subsequent bound-bound decay are observed in
coincidence. Perhaps, the most simple coincidence measurement
refers to the observation of the angular-angular correlations
which reflect the differential alignment in the population of the
magnetic sublevels as function of the observation angle of the
recombination photon. For the electron capture by U$^{\,92+}$
ions, a very strong dependence has been predicted for the angular
distribution of the subsequent photon decay
\cite{Surzhykov/Fri:02,Surzhykov:03b}, using proper tools for the
spin-angular and radial integration
\cite{Fritzsche:97,Surzhykov:04b}. Other correlation functions can
be defined and may lead to additional information about the
polarization properties of the emitted radiation in the future.
The great advantage of such correlation studies is that they
provide an alternative route for determining the polarization
properties of the (heavy) ions beams at storage rings.

\section{Conclusions and outlook}
 \label{outlook}

The photon emission from highly-charged heavy ions has been
reviewed as observed at storage rings. Studies on both, the bound-bound and
free-bound transition in high-$Z$ ions, have revealed many details and have
improved the understanding of the electron-photon interaction in the
strong-fields domain. These investigations clearly show the inherently
\textit{relativistic} behavior of all the structure and collision processes
observed in high-energy atomic physics. In studying high-$Z$ ions,
the role of Quantum Electrodynamics becomes predominant. It is thus
confirmed as the fundamental theory for describing atomic and ionic
systems.

For the capture of electrons into ionic bound states, an
interference has been seen between the different multipole
components in the expansion of the radiation field. These
interference effects have extended our knowledge about the
photoionization of atoms and ions to much higher energies than
available for the neutral elements. They demon\-strate that
magnetic and higher-order contributions of the radiation field may
survive even for energies close to the threshold. Although these
interference effects are observed so far only for the
Lyman-$\alpha$ decay of hydrogen-like uranium, following the
electron capture into the $2p_{3/2}$ level, they are important
also for other few-electron systems if --- apart from the leading
E$_1$ multipole --- other multipole(s) are allowed additionally.
For the emitted photons, then, both the angular distributions and
the polarization properties are likely to be affected.

There are further challenges to be faced in the forthcoming years,
when studying fundamental processes of high-$Z$ ions. For the
radiative recombination, coincidence and polarization measurements
will further advance our knowledge about the electron-photon
interaction as they are sensitive to different components of the
radiation field. As outlined above, moreover, REC may provide a
tool for the diagnostics and detection of spin-polarized ions in
atomic collisions. When compared with accurate theoretical
predictions, the observation of the photon polarization then may
help to control the ion polarization in heavy-storage rings, a
topic which has recently attracted much attention in atomic and
nuclear physics. With control on both, the generation and the
measurement of polarized ion beams, a whole class of new
experiments will become feasible, including the study of parity
non-conserving (PNC) effects \cite{msgi96} or the search for
electric dipole moments of highly-charged ions. Experimentally,
ideal conditions for such challenging studies will be provided by
the new heavy ion facilities presently under discussion, such as
the new international accelerator Facility for Antiproton and Ion
Research (FAIR) at GSI\cite{Henning2001}. There highest
intensities for beams of both stable and exotic heavy nuclei will
become available.
\ack This work has been supported by the BMBF, the GSI and the
Laboratoire Kastler Brossel (Unit{\'e} Mixte de Recherche du CNRS
n$^{\circ}$ 8552).

%
%
%
%
%
\section*{References}
\bibliography{einstein-pi,einstein-sf,einstein-ts}

\end{document}